# Towards Evaluating the Security of Wearable Devices in the Internet of Medical Things

Yas Vaseghi
*Department of Systems and Control*
*Faculty of Electrical Engineering*
*K. N. Toosi University of Technology*
Tehran, Iran
iazzvaseghi@email.kntu.ac.ir

Behnaz Behara
*Department of Biomedical Engineering*
*Faculty of Electrical Engineering*
*K. N. Toosi University of Technology*
Tehran, Iran
b.behara@email.kntu.ac.ir

Mehdi Delrobaei
*Department of Mechatronics*
*Faculty of Electrical Engineering*
*K. N. Toosi University of Technology*
Tehran, Iran
delrobaei@kntu.ac.ir

*Abstract*—The Internet of Medical Things (IoMT) offers a promising solution to improve patient health and reduce human error. Wearable smart infusion pumps that accurately administer medication and integrate with electronic health records are an example of technology that can improve healthcare. They can even alert healthcare professionals or remote servers during operational failure, preventing distressing incidents. However, as the number of connected medical devices increases, the risk of cyber threats also increases. Wearable medication devices based on IoT attached to patients' bodies are prone to significant cyber threats. Being connected to the Internet exposes these devices to potential harm, which could disrupt or degrade device performance and harm patients. To ensure patient safety and well-being, it is crucial to establish secure data authentication for internet-connected medical devices. It is also important to note that the wearability option of such devices might downgrade the computational resources, making them more susceptible to security risks. This paper implements a security approach to a wearable infusion pump. We discuss practical challenges in implementing security-enabled devices and propose initial solutions to mitigate cyber threats.

## I. Introduction

Based on statistics provided by the National Center for Chronic Disease Prevention and Health Promotion (NCCDPHP) [1], 60% of adults in the United States have at least one chronic disease, and 40% suffer from two or more chronic conditions. Many patients with chronic diseases require continuous symptom control and ongoing medication administration. However, frequent visits to healthcare facilities can be challenging for many patients due to the disruptions they cause in their daily lives. Additionally, exceptional circumstances, such as the COVID-19 pandemic, may pose risks to patients and their families, underscoring the need for alternative solutions. In this context, the design and implementation of remote patient monitoring and management systems are essential.

Remote patient monitoring and control systems represent an implementation of smart healthcare solutions. Smart healthcare consists of platforms that connect people, resources, and organizations through wearable devices, IoT, and mobile Internet [2], [3]. Generally, the primary aim of smart healthcare is to ensure timely medical services [4] as chronic diseases continue to escalate and the population ages, which poses significant challenges for the traditional healthcare system. Therefore, to prevent healthcare infrastructures from becoming overwhelmed, in-home telemedicine systems play a pivotal role in the future of healthcare [5].

Our research aims to contribute to advancing smart healthcare through the development of automated and secure medication administration systems. Traditional medication administration approaches often face significant challenges in achieving optimal treatment efficacy [6]. The difficulties lie in accurately administering the correct drug dosage at the most appropriate frequency. As a result, efforts have been directed toward optimizing these parameters [7]. By presenting the development of a smart and secure wearable infusion pump, we seek to take a step further in realizing a secure closed-loop management system for chronic illnesses such as Parkinson's disease [6]. Smart wearable infusion pumps provide accurate dosage and the ability to adapt medication according to real-time vital signs, tailoring treatment to each individual's need [2].

The primary goal of this study lies in implementing a robust security mechanism utilizing tokens for user authentication and fabricating the smart infusion pump. A user-friendly web application interface was developed to facilitate access to patient data by authorized patients and physicians. The interface allows for personalized drug delivery, as drug volume and rate limits can be defined in each patient's profile by their physician.

The server's performance underwent evaluation across three user groups, each varying in the number of users. This assessment centered on two crucial performance metrics: throughput and average response time.

Ultimately, a comprehensive evaluation of the entire system was conducted, involving the measurement of infusion volume and rate, along with the calculation of associated errors.

This paper is organized as follows. Section II provides an overview of prior research in the same field. Section III outlines the methodology, system architecture, and fabricated experimental setup. Section IV showcases the results acquired from the performance evaluations. Lastly, section V provides the conclusion of the study.

## II. Related Work

Zheng *et al.* [8] presented the "Finger-to-Heart" (F2H) authentication system, designed to safeguard wireless implantable medical devices (IMDs). The system leverages a patient's fingerprint for authentication and access permission to the IMDs, emphasizing minimizing resource use in their biometric security approach. In this scheme, the IMD is not required to capture and process biometric data during every access attempt. Additionally, an improved fingerprint authentication algorithm was proposed.

Kulaç [9] introduced a proxy-based protective system designed to be worn externally to enhance the security of wireless implantable medical devices (IMDs) that are not secure. The system ensures secure communication for wireless IMDs by utilizing full-duplex technology. Additionally, sensors incorporated into the jacket provide security at the physical layer.

Jamroz et al. [10] proposed an authentication method for the Internet of Medical Things (IoMD) devices based on hyperelliptic curves and featuring dual signatures. The decreased key size of hyperelliptic curves makes the proposed scheme efficient. Furthermore, the security validation analysis demonstrated the security of the proposed scheme against several types of attacks.

Yu et al. [11] presented a closed-loop glucose-responsive insulin delivery system using a smart insulin patch that delivers insulin in response to increased blood glucose levels. Their research addressed challenges such as fast responsiveness, ease of administration, and biocompatibility.

Wu et al. [12] proposed a self-powered wearable transdermal closed-loop drug delivery system that utilized a hydrogel-based skin patch with side-by-side electrodes for drug delivery. Their system utilized energy harvested from biomechanical motions to facilitate closed-loop drug infusion. The skin patch was designed to enable non-invasive drug infusion through the skin using iontophoresis. This process employs an electric current to drive the flow of charged molecules into the skin.

## III. METHOD

### A. System Architecture

A widely used architecture in healthcare IoT systems adopts a three-tier structure: the sensor or things layer, the communication layer, and the server or processing layer [13]. The system architecture of our designed model also comprises three layers: The application, network, and data storage layers. The application layer includes the physician and an IoT device connected to the patient. An application programming interface (API) was designed as the network layer, an intermediary between the application and data storage layers. Finally, the data storage layer represents the database that stores all the required information for the entire process. Fig. 1 demonstrates the general architecture of the proposed system. A more detailed representation of the implemented model and authorization process is shown in Fig. 2.

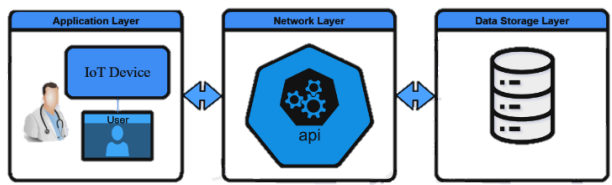

Fig. 1. The general architecture of the proposed system. The application layer consists of a physician and an IoT device. The network layer includes APIs, and the data storage layer represents the database.

*1) Application layer:* This layer is the topmost layer of the system architecture. This layer facilitates communication between the physician and the IoT device. A wearable infusion pump was designed as the IoT device of the proposed architecture. The infusion pump can connect to the Internet and exchange data with the data-storage layer through the designed network layer.

*2) Network layer:* This layer consists of two APIs. It is responsible for connecting the application layer and the data storage layer. It allows the IoT device and the physician to send and receive data to and from the data storage layer. The communication protocol to connect these layers is Hypertext Transfer Protocol (HTTP), which provides a widely supported method for transmitting data over the Internet. The HTTP protocol is a request-response model in which a client sends an HTTP request to a server, which processes the request and returns an HTTP response containing the requested data. The client (application layer) and server communicate through request-response exchanges. The designed APIs and their functionality will be discussed in detail in the next section.

*3) Data storage layer:* Each physician and patient has an account on the developed application. The physician can access the patient's real-time vital data and has the authority to approve or reject drug volume or rate adjustments made by the algorithm. Additionally, all infusions are recorded in the database, allowing the physician to access the patient's infusion history for a comprehensive view of their treatment. Furthermore, each time a request is sent by the infusion pump with an authorized account, the requested information is cached from the data storage layer and sent back to the pump.

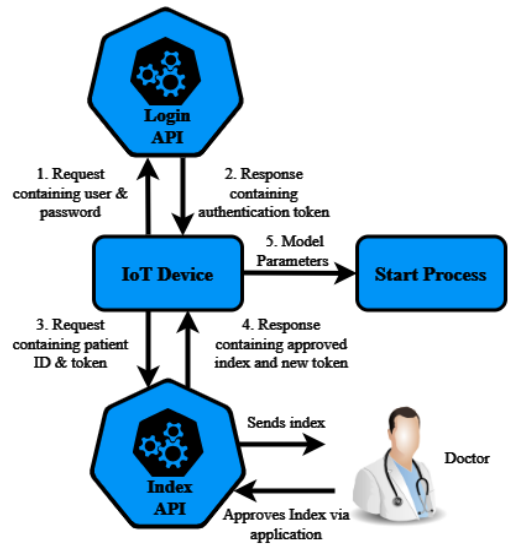

Fig. 2. The detailed representation of the system, which incorporates two APIs. The authorization process utilizes a keyless algorithm by using token-based security.

### B. Experimental Setup

This section presents the design procedure, the hardware components utilized to build the syringe infusion pump, and the software integrated to control its operation while establishing a secure connection to the data storage layer. This compact wearable infusion pump can be worn around the waist or carried in a bag. The performance of the designed infusion pump was tested according to the proposed concept in Fig. 3.

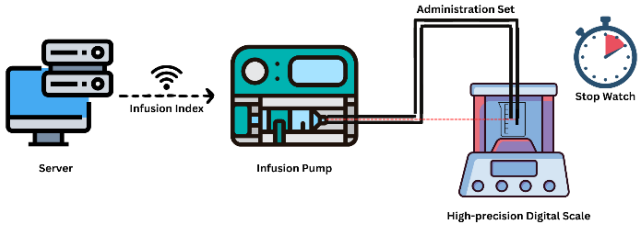

Fig. 3. The diagram of the experimental setup. The infusion index is sent to the pump through the Internet and pump's performance is evaluated by tracking the weight of the medication administered at each instance.

*1) Hardware:* Syringe infusion pumps are electromechanical systems with an accuracy ranging from approximately ±2% to ±5%. These pumps maintain precision and accuracy at low flow rates [14]. Therefore, in this study, the syringe mechanism was adopted as the infusion method for the pump.

An Actuonix P8 miniature linear stepper motor with a precise 0.0018-millimeter full-step size was chosen for precise syringe plunger control in an infusion pump, offering accuracy, compactness (28 grams), and compatibility with low-voltage wearable devices. It operates at a maximum input voltage of 4.2 volts and a maximum current per phase of 256 milliamperes, fitting the energy-efficient requirements of wearables, while its 165:1 gear ratio allows it to handle viscous fluids effectively. Paired with a Pololu DRV8834 low-voltage stepper motor driver offering thermal and current protections, adjustable current control, and a voltage range of 2.5-10.8 volts, the motor is controlled by an ESP32-DevKitC V4, responsible for connectivity, translation of infusion volume to motor steps, and user interface management. The device is powered by a 3.7 V 3400 mAh lithium-ion rechargeable battery with a T6845-C power bank module, ensuring uninterrupted operation even during low battery levels by drawing power from external sources through a micro-USB port.

*2) Software:* Our proposed IoT device is designed to ensure security and precision. We implemented time-expiring one-time-use tokens as a reliable security measure to avoid unauthorized access to the data storage layer through the infusion pump. In the following sections, the authentication and the stepper motor control process will be discussed. Fig. 4 shows the developed web application.

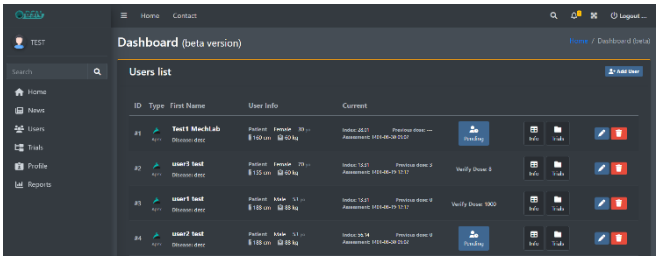

Fig. 4. A screenshot of the developed web application.

*a) Authorization:* As mentioned, our IoT device incorporates two APIs: The login API and the index API. The authorization process incorporates a keyless algorithm by using token-based security. When the user enters their provided username and password into the pump's interface, the microcontroller utilizes these credentials as request parameters. It sends an HTTP POST request to the login API for authentication. Subsequently, the designed application responds to the microcontroller if the provided credentials match the ones on the server. This response comprises four parameters: The user's first name, last name, institute, and a one-time-use time-expiring token. This token acts as a means of authorization to access the resources the index API provides.

Upon receiving the response, the microcontroller extracts the provided token from the response. It then includes this token as a parameter in its new POST request to the index API. Another request utilizes the patient ID as a parameter and the acquired token as a header. If the authentication with the provided token is successful, the index API sends back a response containing two parameters. These parameters include an infusion index and a new one-time-use time-expiring token. The microcontroller extracts these two parameters and initiates the infusion process based on the received index. It also keeps the new token for sending a request to the index API when the infusion is done, and another index is needed. The token provided by the index API is only valid for a set amount of time and expires after that time. If the microcontroller sends a request with an expired token and fails to authenticate, it will send a request to the login API and repeat the process to acquire a new infusion index. Figs. 5 and 6 represent the HTTP message formats of the login and index API, respectively.

Furthermore, each account within the system is linked to the patient's specific device using the device's Media Access Control (MAC) address. The MAC address, a unique identifier assigned to network interfaces, becomes a pivotal security and verification feature, ensuring that users can only access their accounts using a pre-registered and recognized device. This safeguard not only enhances the security of patient data by mitigating unauthorized access but also ensures the integrity and reliability of the information exchanged between the device and the system, as users are precluded from utilizing any device that is not officially registered within their personal account.

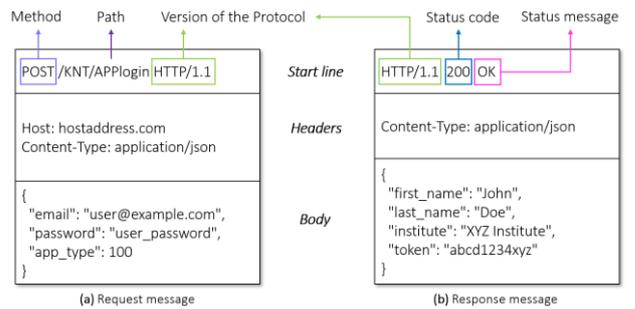

Fig. 5. A representation of the request and response messages (login API).

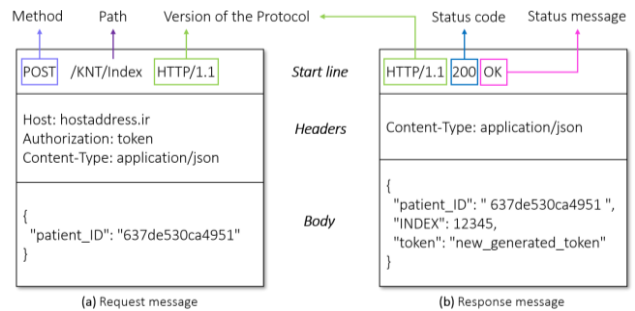

Fig. 6. A representation of the request and response messages (index API).

*b) Polling:* The decision-making algorithm, integral to the operation of the infusion system, may necessitate the modification of the infusion index even during an ongoing infusion process. This is due to various potential reasons, such as a change in the patient's physiological status or an adjustment in medication dosage. Consequently, the infusion pump has been designed to autonomously check its connectivity to the API at predetermined intervals, ensuring continual synchronicity with the latest data.

Upon each connectivity check, the pump reacquires the infusion index. It is imperative to note that if the newly received data presents any discrepancy when compared with the parameters of the ongoing infusion, the system initiates a protocol to modify the delivery parameters. Specifically, new commands are formulated and transmitted to the stepper motor, which is responsible for accurate modulation of the infusion's volume and rate. This adaptive approach ensures that the infusion process remains dynamically aligned with the most updated therapeutic requirements, optimizing patient care and enhancing safety protocols.

Thus, the infusion pump actively collaborates with the algorithm, adapting in real-time and ensuring that the patient receives the most relevant and appropriate medication dosage, making adjustments proficiently and timely, minimizing any risks or delays in administering critical care.

## IV. PERFORMANCE EVALUATION

### A. Server Capability Evaluation

We evaluated the system's performance after designing and implementing the proposed architecture. The evaluation metrics of the proposed system include throughput and response time, which are explained as follows.

- Throughput: the number of requests sent to the server per second.
- Response time: The time of recording patients' data in the system, updating and receiving information between the physician and the patient.

TABLE I. DIFFERENT EVALUATION PARAMETERS FOR THREE USER GROUPS

| User group | Total requests | Avg. response time (ms) | Max. response time (ms) | Min. response time (ms) | Avg. Throughput (requests/sec) |
|---|---|---|---|---|---|
| 20 | 10454 | 62 | 1305 | 42 | 17.02 |
| 50 | 25483 | 66 | 2798 | 41 | 41.32 |
| 100 | 38875 | 184 | 16424 | 42 | 63.21 |

The performance of the proposed architecture was evaluated with three different user groups, including groups of 20, 50, and 100 users. The duration of the performance evaluation was considered 10 minutes. Figs. 7 and 8 show the performance evaluation for 20, 50, and 100 users.

In a time interval of 600 seconds, Fig. 7 shows the average throughput for three user groups. The graph indicates that as the number of users increases, so does the throughput in the system. Similarly, Fig. 8 displays the average response time for the same user groups in the same time interval. The graph shows that the average response time increases as the number of users increases.

Table 1 shows the total number of requests, average throughput, maximum response time, minimum response time, and average response time for three user groups.

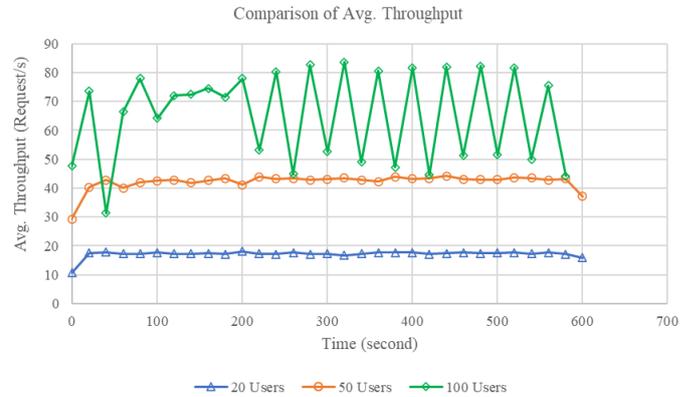

Fig. 7. Comparison of average throughput between three user groups.

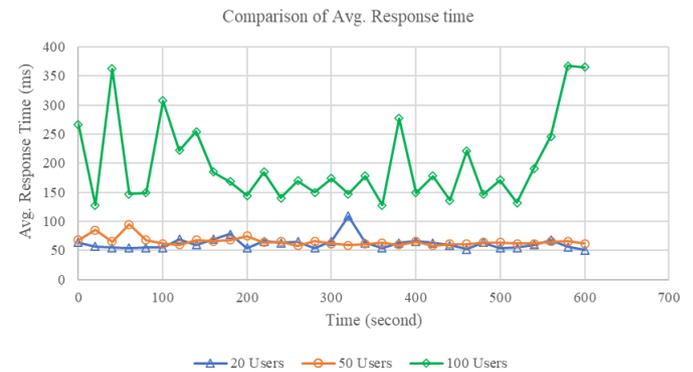

Fig. 8. Comparison of average response time between three user groups.

### B. Infusion Accuracy Evaluation

In order to evaluate the system's performance and accuracy in delivering the desired volume of medication with the desired infusion rate, we employed the setup demonstrated in Fig. 9.

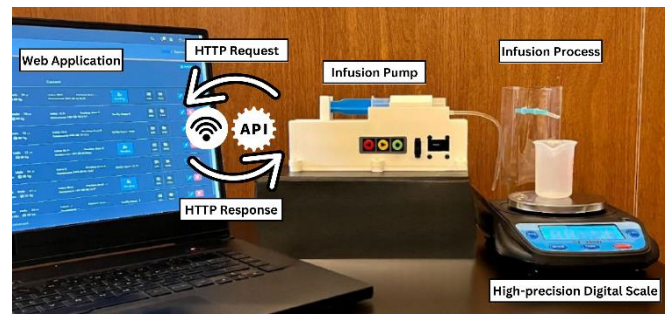

Fig. 9. The experimental setup to evaluate the performance of the system.

The server sends the desired infusion volume and rate to the pump to initiate the infusion process. We used a scale with 0.01 grams accuracy to measure each drop dispensed by the pump. We also started a stopwatch at the beginning of the process to record the exact timing of each drop as it was infused. This data helped us calculate the volume of each drop infused and the following infusion rate.

This procedure was repeated 10 times with two different infusion settings. The first setting, consisting of a desired volume of 2 milliliters and a rate of 4 milliliters per hour, was repeated five times. The second setting, with a desired volume of 5 milliliters and a rate of 5 milliliters per hour, was also repeated five times.

Fig. 10 demonstrates the infused volume over time for all five experiments of the first setting.

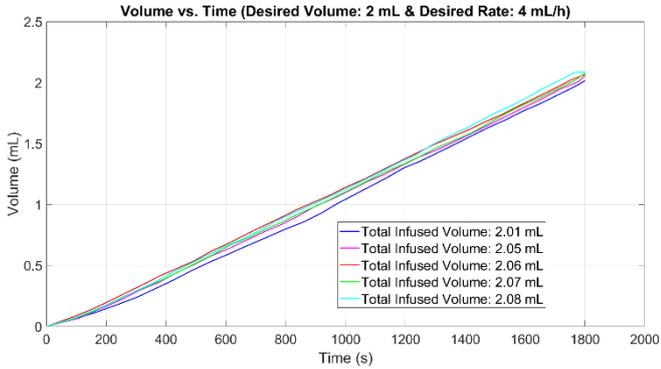

Fig. 10. Infused volume over time for all five experiments of the first setting (desired volume: 2 ml, desired rate: 4 ml/h, infusion time: 1800 s).

Fig. 11 demonstrates the infusion rate over time for all five experiments of the first setting.

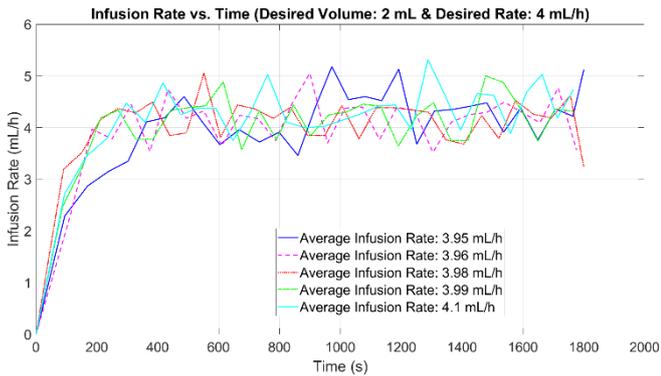

Fig. 11. Infusion rate over time for all five experiments of the first setting (desired volume: 2 ml, desired rate: 4 ml/h, infusion duration: 1800 s).

Fig. 12 demonstrates the infused volume over time for all five experiments of the second setting.

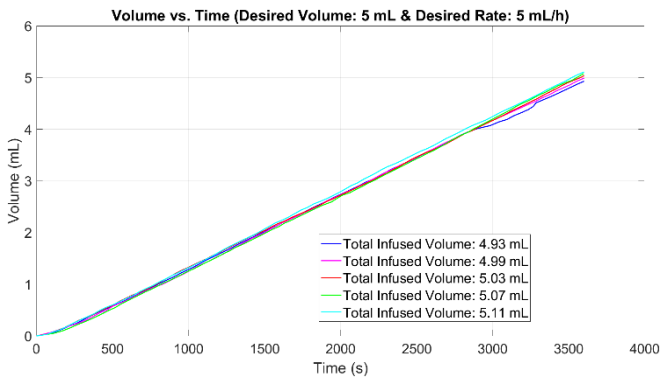

Fig. 12. Infused volume over time for all five experiments of the second setting (desired volume: 5 ml, desired rate: 5 ml/h, infusion duration: 3600 s).

Fig. 13 demonstrates the infusion rate over time for all five experiments of the second setting.

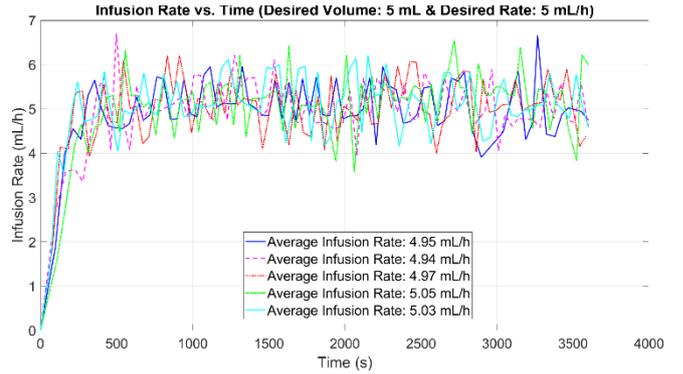

Fig. 13. Infusion rate over time for all five experiments of the second setting (desired volume: 5 ml and desired rate: 5 ml/h, infusion duration: 3600 s).

The final results of the performed experiments have been summarized in Table II.

TABLE II. SYSTEM EVALUATION RESULTS

| Setting | Experiment | Delivered Volume (mL) | %Error in Volume | Average Infusion Rate (mL/h) | %Error in Infusion Rate |
|---|---|---|---|---|---|
| Volume = 2 mL Rate = 4 mL/h | 1 | 2.05 | 2.5% | 3.96 | 1% |
| | 2 | 2.07 | 3.5% | 3.99 | 0.25% |
| | 3 | 2.01 | 0.5% | 3.95 | 1.25% |
| | 4 | 2.08 | 4% | 4.1 | 2.5% |
| | 5 | 2.06 | 3% | 3.98 | 0.5% |
| | Avg. %Error in Volume | | 2.7% | Avg. %Error in Infusion Rate | 1.1% |
| Volume = 5 mL Rate = 5 mL/h | 1 | 5.03 | 0.6% | 4.97 | 0.6% |
| | 2 | 4.93 | 1.4% | 4.95 | 1% |
| | 3 | 5.07 | 1.4% | 5.05 | 1% |
| | 4 | 5.11 | 2.2% | 5.03 | 0.6% |
| | 5 | 4.99 | 0.2% | 4.94 | 1.2% |
| | Avg. %Error in Volume | | 1.16% | Avg. %Error in Infusion Rate | 0.88% |

## V. Conclusion

This paper detailed a data authentication process to ensure the safety of an infusion pump that can be accessed and managed remotely. The primary objective of this work was to focus on patient care for chronic illnesses by implementing a secure approach for the IoT-enabled wearable infusion pumps. The system's multi-layered architecture facilitated efficient communication in the proposed configuration, while the web interface allowed authorized patients and physicians to customize drug delivery settings. Security was mainly ensured by generating unique tokens that expire immediately after the first use, and each account is tied to a specific device through its MAC address. Moreover, the infusion pump adjusts autonomously to changing therapeutic requirements to ensure prompt and safe medication administration. The system's server performance assessment indicated that it can handle multiple user groups, and the overall design was found to be reliable and practical.